\documentclass[11pt]{article}
\pdfoutput=1
\usepackage{jcappub,cancel}
\usepackage{paralist}
\usepackage{booktabs}
\usepackage[a4paper]{geometry}
\usepackage{hyperref}
\usepackage{xspace}

\newcommand{\GeV}{\mathrm{GeV}}

\newcommand{\beq}{\begin{equation}}
\newcommand{\eeq}{\end{equation}}

\newcommand{\sarah}{\texttt{SARAH}\xspace}

\newcommand{\cosmotransitions}{\texttt{CosmoTransitions}\xspace}

\newcommand{\gev}{\, {\rm GeV}}

\newcommand{\eg}{\emph{e.g.},\xspace}
\newcommand{\ie}{\emph{i.e.},\xspace}
\newcommand{\Ztwo}{\mathbb{Z}_2}

\title{The Electroweak Phase Transition in the Inert Doublet Model}

\author[a,b,c,d]{Nikita Blinov,}
\author[a,b]{Stefano Profumo,}
\author[a,b]{Tim Stefaniak}
\affiliation[a]{Department of Physics, University of California, Santa Cruz\\1156 High St, Santa Cruz, CA 95064}
\affiliation[b]{Santa Cruz Institute for Particle Physics,\\ 1156 High St, Santa Cruz, CA 95064}
\affiliation[c]{Theory Department, TRIUMF, 4004 Wesbrook Mall, Vancouver, BC V6T 2A3, Canada}
\affiliation[d]{Department of Physics and Astronomy, University of British Columbia, Vancouver, BC V6T 1Z1, Canada}
\emailAdd{nblinov@triumf.ca}
\emailAdd{profumo@ucsc.edu}
\emailAdd{tistefan@ucsc.edu}

\abstract{We study the strength of a first-order electroweak phase transition
in the Inert Doublet Model (IDM), where particle dark matter (DM) is comprised of the lightest neutral inert Higgs boson. We improve over previous studies in the
description and treatment of the finite-temperature effective potential and of
the electroweak phase transition. We focus on a set of benchmark models
inspired by the key mechanisms in the IDM leading to a viable dark matter
particle candidate, and illustrate how to enhance the strength of the
electroweak phase transition by adjusting the masses of the yet undiscovered
IDM Higgs states. We argue that across a variety of DM masses, obtaining a
strong enough first-order phase transition is a generic possibility in the IDM.
We find that due to direct dark matter searches and collider
constraints, a sufficiently strong transition and a thermal relic density
matching the universal DM abundance is possible only in the Higgs funnel
regime.}

\begin{document}

\maketitle

\section{Introduction}

The simplest extension of the Standard Model (SM) that includes two $SU(2)$
Higgs doublets is known as the inert Higgs Doublet Model (IDM). In the IDM, the
extra doublet has no coupling to SM fermions and is odd under a postulated new
$\Ztwo$ discrete symmetry, whereas all SM fields are $\Ztwo$-even. 
Such symmetry makes the lightest $\Ztwo$-odd particle (LOP) from the
extra doublet stable and, thus, a potential weakly interacting massive particle
(WIMP) dark matter candidate. The symmetry also eliminates 
numerous terms in the interaction Lagrangian of the model containing an odd
number of extra ``inert'' scalars.

The IDM was introduced originally as a possible generic scenario for
electroweak symmetry breaking (EWSB) \cite{Deshpande:1977rw}. Only subsequently
was it realized that the IDM naturally features a WIMP DM
candidate~\cite{Ma:2006km, Barbieri:2006dq}, possibly providing a thermal relic
density compatible with the inferred universal DM abundance. Numerous studies
have subsequently investigated the DM and collider phenomenology of the model
(see, \eg~Ref.~\cite{Gustafsson:2007pc, Agrawal:2008xz, Andreas:2009hj,
Nezri:2009jd,Arina:2009um, Gong:2012ri, Gustafsson:2012aj}).

An additional early motivation to consider the IDM as an appealing augmentation
of the SM scalar structure was to allow for a relatively heavy SM-like Higgs
while remaining compatible with constraints from electroweak precision
observables, and without large fine tuning
\cite{Barbieri:2006dq,Gustafsson:2012aj}. Although this motivation has
somewhat faded after the discovery of a SM-like Higgs boson at the LHC with a
mass of $\sim 125~\GeV$~\cite{Aad:2012tfa,Chatrchyan:2012ufa}, this important discovery decreases the number of free
parameters in the theory by one, and places interesting and stringent
constraints on the IDM phenomenology~\cite{Goudelis:2013uca,Khan:2015ipa}. 

In the present study we are concerned with the nature of the electroweak phase
transition (EWPT) in the IDM, and, specifically, with determining which
physical parameters drive the strength of the phase transition, making it more
or less strongly first-order, or second-order. This question is intimately
related with the possibility to produce the observed baryon-antibaryon
asymmetry in the Universe at the electroweak phase transition: A strongly
first-order phase transition (in a quantitative sense we shall make clear
below) is a necessary ingredient to {\em (i)} achieve the necessary
out-of-equilibrium conditions, occurring on the boundary of broken and unbroken
electroweak phase, and to {\em (ii)} shield a baryon asymmetry captured in the
broken electroweak phase region from sphaleron wash-out.

While necessary, a strongly first-order phase transition is not a sufficient
condition. The $CP$ violating sources of the SM are known to be insufficient to
generate the necessary asymmetry in the number density of baryons compared to
antibaryons during the electroweak phase transition. The unbroken $\Ztwo$
symmetry in the IDM precludes any new source of $CP$ violation, and thus this
model per se cannot accommodate successful electroweak baryogenesis (EWBG).
However, the IDM might be in effect a good approximation at low energy of a
broader construction that includes such additional $CP$ violating sources at
higher energies. Several suggestions of plausible effective higher-dimensional
operators have been made in the literature \cite{Dine:1990fj, Dine:1991ck}. We
will not discuss this aspect any more, as it falls outside the scope of this
study.

The nature and strength of the electroweak phase transition in the IDM has been
subject of several studies, with increasingly refined treatment of the
effective potential~\cite{Chowdhury:2011ga,
Borah:2012pu,Gil:2012ya,AbdusSalam:2013eya,Cline:2013bln}. For example,
Ref.~\cite{Chowdhury:2011ga} utilized only the high-temperature form of the
effective potential without including the zero-temperature Coleman-Weinberg
terms. These were then shown to be quantitatively important for the phase
transition strength in Ref.~\cite{Gil:2012ya}, where the full one-loop
effective potential was used. Alternative $SU(2)_L$ representations of the
inert scalar were considered in Ref.~\cite{AbdusSalam:2013eya} where it was
argued that in general, higher representations are less successful in
satisfying experimental and theoretical constraints, thereby further motivating
the study of the doublet case. 

With the exception of Ref.~\cite{Cline:2013bln}, the primary focus has been 
 on the Higgs funnel regime (described in more detail in Section~\ref{sec:idm_dm}). 
Indeed, we will confirm the findings of Refs.~\cite{Borah:2012pu,Gil:2012ya} 
that this is the only region of parameter space that 
can successfully saturate the DM abundance and provide a strong-enough first-order 
EWPT. In Ref.~\cite{Cline:2013bln} it was emphasized that the IDM 
can be useful for the EWPT even if the LOP provides only a sub-leading 
component of DM.

In this work we go beyond previous studies by utilizing a state-of-the-art
treatment of the finite-temperature and zero-temperature effective potential
including renormalization group, daisy resummation improvements and one-loop
model parameter determination. As we discuss in great detail in what follows, strongly
first-order EWPT in the IDM requires sizable quartic couplings that enhance
quantum corrections to masses. This is important in the context of DM
phenomenology since DM particle production in the early Universe often relies
on resonance and threshold effects~\cite{Goudelis:2013uca}. In addition, we
also ensure that the phase transition completes by evaluating bubble nucleation
rates. 

Unlike previous studies which primarily utilized large numerical scans of the
parameter space, here we take an orthogonal approach: we restrict our attention
to a few benchmark models, motivated by the requirement of having a viable
dark matter particle candidate and representing different features in the DM
phenomenology. Based on these benchmarks, we then discuss how the EWPT depends
on the physical model parameters. We identify the key physical inputs that
drive the phase transition to the interesting regime where it is strongly
enough first-order to accommodate successful electroweak baryogenesis. We will
see that in all but one case the demand for a strongly first-order EWPT is in
tension with either the relic abundance requirement or with experimental
probes.

Our central finding is that the main driver of the strength of the phase
transition is the mass difference between the lightest inert scalar and the
heavier scalars. Thus, we extend the results of Refs.~\cite{Chowdhury:2011ga,Gil:2012ya} 
to other regions of IDM parameter space. For large enough mass splittings, but for light enough heavy
scalars, we find a phase transition strength (as measured by the ratio
$v_c/T_c$, as we discuss in detail below) which increases with the
mass splitting.

The remainder of this paper is organized as follows: In
Section~\ref{sec:pt_in_idm} we give a brief introduction to the IDM, thereby
clarifying our conventions, discuss quantum and finite-temperature corrections
to the effective potential, and outline 
the computation of the phase transition strength. The essential features 
of DM phenomenology are reviewed in Section~\ref{sec:idm_dm}. 
In Section~\ref{sec:benchmarks} we study the electroweak 
phase transition in several benchmark models motivated by the various 
DM scenarios available in the IDM. We conclude in Section~\ref{sec:conclusion}.

\section{Phase Transitions in the Inert Doublet Model (IDM)\label{sec:pt_in_idm}}
\subsection{IDM at Tree-Level}
The IDM is a particular realization of the general type I Two Higgs Doublet
Model (2HDM) (see, \eg, Ref.~\cite{Branco:2011iw} for a review) which features
an additional $\Ztwo$ symmetry. The SM doublet $H$ is even under $\Ztwo$, while
the new (inert) doublet $\Phi$ is odd. If we take $\Phi$ to have hypercharge
$+1/2$, the most general renormalizable potential consistent with these
symmetries is then given by~\cite{Goudelis:2013uca}:
\beq
V_0 = \mu_1^2 |H|^2 + \mu_2^2|\Phi|^2 + \lambda_1 |H|^4+ \lambda_2 |\Phi|^4 
+ \lambda_3 |H|^2| \Phi|^2 + \lambda_4 |H^\dagger\Phi|^2 + 
\frac{\lambda_5}{2} \left[ (H^\dagger\Phi)^2 + \mathrm{h.c.} \right].
\label{eq:treepot}
\eeq

Conventionally, within $CP$-conserving Higgs sectors, the physical states are
decomposed into $CP$-even and $CP$-odd scalars. One should keep in mind,
however, that in the IDM there is no observable that can actually distinguish
between the $CP$-even or $CP$-odd character of the inert Higgs bosons. In the
absence of a vacuum expectation value (VEV) for $\Phi$, the doublets decompose
as 
\beq
H = \left( \begin{array}{c} G^+ \\ \frac{1}{\sqrt{2}}\left(v+ h+\mathrm{i}G^0\right)
 \end{array} \right),
	\qquad
\Phi = \left( \begin{array}{c} H^+\\ \frac{1}{\sqrt{2}}\left(H +\mathrm{i}A \right)
 \end{array} \right). 
\eeq
Below we consider the thermal evolution of the effective scalar potential
in the early Universe. In general, spontaneous breaking of $\Ztwo$ can occur,
in which case we must also include a VEV for the neutral component of $\Phi$
which we will indicate with $\phi$. 

The lightest $\Ztwo$-odd particle is stable, and potentially
provides a viable particle dark matter candidate. The $\Ztwo$ symmetry also
forbids Yukawa couplings of $\Phi$ to SM fermions (assumed to be
even under $\Ztwo$), 
which eliminates tree-level flavor-changing neutral currents. Either $H$ or $A$
can be the LOP, and since gauge interactions with SM states do not distinguish
between the two, they are effectively equivalent from the standpoint of
phenomenology. Below we will indicate the LOP as $H$, but all statements made
with respect to DM phenomenology and the electroweak phase transition remain
true after the replacements $H\rightarrow A$ and $\lambda_L\rightarrow\lambda_S$, 
where $\lambda_L = (\lambda_3 + \lambda_4 + \lambda_5)/2$ 
and $\lambda_S = (\lambda_3 + \lambda_4 - \lambda_5)/2$ determine the coupling 
of the LOP to the SM Higgs~\cite{Borah:2012pu}.

In the electroweak vacuum, the tree-level masses of the new states are given by 
\begin{align}
m_h^2 & = \mu_1^2 + 3\lambda_1 v^2, \nonumber \\
m_H^2 & = \mu_2^2 + \lambda_L v^2,\nonumber \\
m_A^2 & = \mu_2^2 + \lambda_S v^2,\nonumber \\
m_{H^\pm}^2 & = \mu_2^2 + \frac{1}{2}\lambda_3 v^2.\label{eq:treemasses}
\end{align}
The determination of model parameters from physical inputs is discussed in the 
next section. 

\subsection{Finite-Temperature Corrections\label{sec:ft_corr}}
The effective potential at finite temperature $T$ can be written as
\beq
V_\mathrm{eff} = V_0 + V_1 + V_T,
\eeq
where $V_0$, $V_1$ and $V_T$ are tree-level, one-loop temperature-independent
and -dependent pieces, respectively. The tree-level potential $V_0$ has been
given in Eq.~\eqref{eq:treepot}. Working in the Landau gauge ($\xi=0$), the temperature-independent one-loop
correction has the Coleman-Weinberg form~\cite{Brandenberger:1984cz,
Sher:1988mj, Quiros:1999jp}:
\beq
V_1 = \sum_i \frac{n_i}{64\pi^2} m^4_i(v,\phi) \left(\ln \frac{m^2_i(v,\phi)} {Q^2} - C_i \right).
\eeq
The sum is over all particle species coupling to the doublets;
 $n_i$ is the number of degrees of freedom (positive for bosons and 
negative for fermions), $C_i$ are renormalization-scheme-dependent constants 
($C_i = 1/2$ for transverse gauge bosons and $3/2$ for 
everything else in the $\overline{\mathrm{MS}}$ scheme); $m^2_i(v,\phi)$ 
is the field-dependent squared mass 
for each species. In writing the above, we have implicitly absorbed the 
counterterms into $V_1$; the temperature-dependent part is ultraviolet finite.
The counterterms and, equivalently, the renormalized parameters $\mu_1^2$ and $\lambda_1$ are 
determined by ensuring that the one-loop potential reproduces the physical values $v=246.22\gev$ and $m_h\approx125\gev$. 
The remaining parameters of the model are specified using the three physical masses $m_H$, $m_A$ 
and $m_{H^\pm}$, along with $\lambda_L$ and $\lambda_2$. The masses are related 
to potential parameters using the one-loop relations from Ref.~\cite{Goudelis:2013uca}, while 
$\lambda_L$ and $\lambda_2$ are taken to be running $\overline{\mathrm{MS}}$ values defined at scale $M_Z$.
These requirements specify the renormalization conditions. 

The field dependent masses in the IDM for the SM vector bosons and fermions are, respectively,
\beq
m_W^2 = \frac{1}{4} g^2 (v^2 + \phi^2) \label{eq:fdWmass},\quad
m_Z^2 = \frac{1}{4} (g^2 + g'^2) (v^2 + \phi^2),\quad m_\gamma^2 = 0 
\eeq
and
\beq
m_f^2 = \frac{1}{2}y_f^2 v^2,
\eeq
with the corresponding bosonic degrees of freedom $n_i=6$, $3$, $2$ for $i=W$, $Z$, $A$, and fermionic degrees of freedom 
$n_i=-12$, $-12$, $-4$ for $i=t$, $b$, $\tau$.
 
The field-dependent neutral $CP$-even, $CP$-odd and charged scalar mass
eigenstates are obtained by diagonalizing 
\begin{align}
M^2_h & = \begin{pmatrix}
\mu_1^2 + 3\lambda_1 v^2+ \lambda_L \phi^2 & 2\lambda_L \phi v \\
 2\lambda_L \phi v & \mu_2^2 + 3\lambda_2 \phi^2 + \lambda_L v^2
\end{pmatrix} \label{eq:CPEvenMass}\\
M^2_A & = \begin{pmatrix}
\mu_1^2 + \lambda_1 v^2+ \lambda_S \phi^2 & \lambda_5 \phi v \\
 \lambda_5 \phi v & \mu_2^2 + \lambda_2 \phi^2 + \lambda_S v^2
\end{pmatrix} \label{eq:CPOddMass}\\
M^2_\pm &=\begin{pmatrix}
\mu_1^2 + \lambda_1 v^2+ \frac{1}{2}\lambda_3 \phi^2 & \frac{1}{2}(\lambda_5+\lambda_4) \phi v \\
\frac{1}{2}(\lambda_5+\lambda_4) \phi v & \mu_2^2 + \lambda_2 \phi^2 + \frac{1}{2}\lambda_3 v^2
\end{pmatrix}.
\label{eq:ChargedMass}
\end{align}
Equations~\eqref{eq:CPOddMass} and~\eqref{eq:ChargedMass} include contributions both from physical states and 
from Goldstone bosons. Notice that for $\phi=0$ the $(22)$ components reduce to the expressions in
Eq.~\eqref{eq:treemasses}. 

The leading order quantum corrections give rise to a 
renormalization scale-dependent potential. One can choose
the renormalization scale $Q$ to minimize the size of higher order $k$-loop 
corrections which scale with $(\ln m^2 /Q^2)^k$. The scale choice 
can be important when a parameter in the potential is very different 
from the electroweak VEV $\sim 246\;\GeV$. We thus choose to use the renormalization 
group (RG) improved effective potential to minimize the scale dependence. The 
potential parameters are replaced by their running values, evaluated 
at the scale $Q$. The relevant one-loop $\beta$ functions are 
given in Appendix~\ref{sec:rge}. Our computations are performed with $Q=246\;\GeV$. 

The leading order temperature-dependent corrections to the effective potential in the Landau gauge 
take the form~\cite{Quiros:1999jp} 
\beq
V_T = 
\frac{T^4}{2\pi^2}\left(\sum_{i=\mathrm{bosons}} n_i J_B\left[m^2_i(v,\phi)/T^2\right]
+ 
\sum_{i=\mathrm{fermions}} n_i J_F\left[m^2_i(v,\phi)/T^2\right]
\right),
\eeq
where the $J$ functions are defined as 
\begin{align}
J_B(x) & = \int_0^\infty dt\; t^2 \ln\left[1 - \exp\left(-\sqrt{t^2 + x}\right)\right],\label{eq:JB}\\
J_F(x) & = \int_0^\infty dt\; t^2 \ln\left[1 + \exp\left(-\sqrt{t^2 + x}\right)\right].
\end{align}
These functions admit useful high-temperature expansions which allow us to 
study the phase structure as a function of $T$ analytically (as long as the expansion is 
justified):
\begin{align}
T^4 J_B\left[m^2/T^2\right] & = -\frac{\pi^4 T^4}{45}+
\frac{\pi^2}{12} T^2 m^2-\frac{\pi}{6}
T \left(m^2\right)^{3/2}-\frac{1}{32} 
m^4\ln\frac{m^2}{a_b T^2} + \mathcal{O}\left(m^2/T^2\right)\label{eq:Jb_highT}\\ 
T^4 J_F\left[m^2/T^2\right]& = \frac{7\pi^4 T^4}{360}-
\frac{\pi^2}{24} T^2 m^2-\frac{1}{32}
m^4\ln\frac{m^2}{a_f T^2} + \mathcal{O}\left(m^2/T^2\right),\label{eq:Jf_highT} 
\end{align}
where $a_b = 16 a_f = 16\pi^2 \exp(3/2 - 2\gamma_E)$. The $T^2$ terms in the
expressions above illustrate symmetry restoration at high temperatures. The
non-analytic $m^3$ term in Eq.~\eqref{eq:Jb_highT} can be responsible for the
barrier between the high $T$ phase (at the field origin) and low $T$ phase that
breaks $SU(2)_L \times U(1)_Y$.

Note that symmetry restoration signals the breakdown of perturbation theory --- higher order 
diagrams become important. This can be accounted for by performing a resummation of 
daisy diagrams~\cite{Parwani:1991gq, Arnold:1992rz, Espinosa:1995se}. 
The resummation is performed by adding finite-temperature corrections to the
boson masses in Eq.~\eqref{eq:JB}: 
\beq
m^2 \rightarrow m^2 + c T^2, 
\eeq
where $c$ is computed from the infrared limit of the corresponding two-point function. 
For the SM Higgs doublet we find 
\beq
c_1 = \frac{1}{8} g^2 + \frac{1}{16}(g^2 + g^{\prime 2}) + \frac{1}{2}\lambda_1 
+ \frac{1}{12}\lambda_L + \frac{1}{12}\lambda_S+ \frac{1}{12}\lambda_3
+ \frac{1}{4} y_t^2 + \frac{1}{4} y_b^2 + \frac{1}{12} y_\tau^2.
\eeq 
The various components of the inert doublet receive similar contributions (but 
without contributions from the fermions):
\beq
c_2 = \frac{1}{8} g^2 + \frac{1}{16}(g^2 + g^{\prime 2}) + \frac{1}{2}\lambda_2 
+ \frac{1}{12}\lambda_L + \frac{1}{12}\lambda_S+ \frac{1}{12}\lambda_3.
\eeq 
These expressions are in agreement with those in Refs.~\cite{Carrington:1991hz,Cline:1995dg,Gil:2012ya}. 
We implement these corrections by replacing $\mu_i^2 \rightarrow \mu_i^2 + c_i T^2$ in 
the scalar mass matrices, Eqs.~(\ref{eq:CPEvenMass}, \ref{eq:CPOddMass}, \ref{eq:ChargedMass}).\footnote{There are subleading thermal corrections to off-diagonal self-energies suppressed by 
additional powers of coupling constants and VEVs which are usually neglected.}

The thermal masses of the gauge bosons are more complicated. 
Only the longitudinal components receive corrections. The expressions for 
these in the SM can be found in Ref.~\cite{Carrington:1991hz}, but it is easy to modify 
them to include the contribution of an extra Higgs doublet. For the longitudinally polarized $W$ boson, 
the result is 
\beq
m_{W_L}^2 = m_W^2 + 2 g^2 T^2.
\eeq 
This includes contributions from gauge boson self-interactions, two Higgs doublets and 
all three fermion families. 
The masses of the longitudinal $Z$ and $A$ are determined by diagonalizing the matrix 
\beq
\frac{1}{4}(v^2 + \phi^2)
\begin{pmatrix}
g^2& -g g^\prime \\
-g g^\prime & g^{\prime 2}
\end{pmatrix}
+
\begin{pmatrix}
2 g^2 T^2 & 0 \\
0 & 2 g^{\prime 2} T^2 
\end{pmatrix}.
\eeq
The eigenvalues can be written as
\beq
m_{Z_L,A_L}^2 = \frac{1}{2} m_Z^2 + (g^2 + g^{\prime 2} )T^2 \pm \Delta, 
\eeq
where 
\beq
\Delta^2 = \left(\frac{1}{2} m_Z^2 + (g^2 + g^{\prime 2} )T^2 \right)^2
- g^2 g^{\prime 2} T^2 ( v^2 + \phi^2 + 4 T^2). 
\eeq
\subsection{Electroweak Phase Transition (EWPT)}
Armed with the finite-temperature effective potential, we now proceed to study the
structure of the EWPT. The key property we intend to investigate is the
transition strength, which sets the baryon number wash-out rate inside a bubble
of broken phase (for a recent review of electroweak baryogenesis, see, \eg
Ref.~\cite{Morrissey:2012db}). In order to suppress sphaleron wash-out in the
regions of broken electroweak phase, the relevant condition is typically
quantified by requiring that \cite{Kuzmin:1985mm}
\beq
\frac{v_c}{T_c} \gtrsim 1, 
\label{eq:bnpc_gauge_variant}
\eeq
where $v_c$ is the Higgs VEV at the critical temperature $T_c$, defined as the
temperature at which the origin is degenerate with the electroweak-breaking
vacuum. 

Note that it has been shown that this baryon number preservation condition
(BNPC) is a quantity which is manifestly not
gauge-invariant~\cite{Patel:2011th}. A gauge invariant BNPC can be however
derived from the high-$T$ expansion of the dimensionally reduced effective
action and the critical temperature $T_c$ must be obtained using the gauge
invariant prescription of Ref.~\cite{Patel:2011th}, which employs expansions in
powers of $\hbar$ of the potential and VEV. Near the critical
temperature, $\mathcal{O}(\hbar)$ contributions to the potential are as
important as the tree-level terms, so the $\hbar$ expansion fails. This is
also why an all-orders ring diagram resummation discussed in
Sec.~\ref{sec:ft_corr} is needed. A consistent gauge-invariant method for implementing 
the ring resummation for the effective potential evaluated at the minimum 
was also demonstrated in Ref.~\cite{Patel:2011th}. We will be interested in studying 
tunneling and nucleation temperatures, which require the evaluation of the potential 
away from the minima. For this reason, below we employ the
standard BNPC of Eq.~\eqref{eq:bnpc_gauge_variant} and use the full one-loop
effective potential to study IDM phases. We will argue that our results do not
depend strongly on the issues of gauge invariance. We leave the 
full gauge-invariant treatment of the IDM to future work.

Finally let us note that the physical phase transition does not begin at $T_c$, but rather 
at a lower \emph{nucleation} temperature $T_n$, at which the bubble formation rate exceeds 
the Hubble expansion rate. If this rate is too slow, the false vacuum is metastable 
and the transition does not complete. We evaluate the nucleation 
temperature for a given model with a first-order phase transition using the 
\cosmotransitions\ package~\cite{Wainwright:2011kj}.

\section{Dark Matter\label{sec:idm_dm}}

The requirement of a thermal relic abundance for the LOP matching the observed DM 
density in the Universe of $\Omega_{\mathrm{cdm}}h^2 = 0.1199\pm0.0022$~\cite{Planck:2015xua}, or 
at least of not over-producing such density via thermal production (``subdominant IDM'', 
see, \eg Ref.~\cite{Cline:2013bln}) naturally selects four distinct 
sectors of the model's parameter space:
\begin{enumerate}
\item a \emph{low} mass regime, with a LOP mass, $m_H$, well 
below half the observed SM-like Higgs mass, $m_H\lesssim m_h/2$;
\item a \emph{resonant} or \emph{funnel}~region, $m_H\sim m_h/2$, \ie a mass range where 
LOP annihilation proceeds predominantly through quasi on-shell Higgs $s$-channel exchange;
\item an \emph{intermediate} mass regime, with a LOP mass of $m_h/2\ll m_H\lesssim 500\gev$;
\item a \emph{heavy} mass regime, with a LOP mass between $500\gev$ a few TeV.
\end{enumerate}

In the first case, the \emph{low} mass regime, the DM pair-annihilation
predominantly proceeds via the pair production of the heaviest kinematically
accessible fermion ($\tau$ leptons, $b$ quarks) through $h$ exchange. The
lower the LOP mass, the larger the $\lambda_{L,S}$ couplings need to be in
order to produce a large enough pair-annihilation cross section. The allowed
mass values range down to values close to the classical Lee-Weinberg lower mass
limit for WIMPs~\cite{Lee:1977ua}, for this class of models somewhere in
between 3 and 4 GeV. Direct detection limits from XENON10 \cite{Angle:2011th}
probe such combinations of masses and couplings quite tightly, such that only a
small mass window below $5-7\gev$ remains.\footnote{The exact limit depends on
the different possible choices of nucleon matrix elements, especially those connected
with the strange quark content of nucleons \cite{Young:2013nn}.}

As the mass of the LOP approaches the resonant condition $m_H\sim m_h/2$, the
resonant Higgs exchange allows for much smaller values of the $\lambda_{L,S}$
couplings, and direct detection constraints can be readily evaded. The
relevant mass window left unconstrained by~XENON100~\cite{Aprile:2012nq} and
LUX~\cite{Akerib:2013tjd} has a width of approximately $10-15\gev$ centered
around $m_h/2$~\cite{Arhrib:2013ela}.

The mass regions above and below the resonance $m_H = m_h/2$ are actually
slightly different from each other: Above the resonance, the pair production of
$WW^{*}$ in the final state of DM pair annihilation processes becomes increasingly important, even if
$\lambda_{L,S}=0$, because the four-point interaction through gauge couplings,
independent of $\lambda_{L,S}$, starts contributing significantly. As a result
the values of $\lambda_{L,S}$ giving the ``correct'' relic density are pushed
to increasingly (with LOP mass) large, negative values. 

For larger and larger LOP masses, the cross section for 
LOP pair annihilation to gauge bosons becomes very large, such that the thermal relic density is 
systematically below the universal dark matter density for any combination 
of model parameters. Barring non-thermal production mechanisms, in this 
intermediate mass region the LOP cannot be the dominant dark 
matter constituent~\cite{Goudelis:2013uca, Cline:2013bln}.

Finally, at about $m_H\simeq500~\GeV$, for $\lambda_{L,S}\simeq0$ cancellations 
between scalar $t$- and $u$-channel exchange diagrams and the four-point 
interaction diagram alluded to above allow, again, for a sufficiently large thermal relic density. 
Such cancellations are suppressed by driving $\lambda_{L,S}$ away from zero. 
Thus, tuning $\lambda_{L,S}$ for increasing values of $m_H$ generally allows one 
to achieve the correct relic density 
for mass values from $m_H \gtrsim 500\gev$ up into the multi-TeV range. 
This heavy mass regime of the IDM can be seen as a low energy 
effective theory of a composite dark sector~\cite{Carmona:2015haa}.

\section{Benchmark Models\label{sec:benchmarks}}
The benchmark models specified in Ref.~\cite{Goudelis:2013uca} demonstrated
various aspects of DM phenomenology and the possibility for the IDM to
influence the $h\rightarrow\gamma\gamma$ rate. Unfortunately, none of the
suggested scenarios exhibits a strongly first-order EW phase transition. In
this Section, we identify alternate benchmark models which can potentially
yield a strongly first-order EW phase transition, while having disparate
properties for the lightest $\Ztwo$-odd particle. All our benchmark
models are compatible with constraints from Higgs collider bounds and rate
measurements, which has been explicitly checked using the tools
\texttt{HiggsBounds}~\cite{Bechtle:2008jh,Bechtle:2011sb,Bechtle:2013wla} and
\texttt{HiggsSignals}~\cite{Bechtle:2013xfa}, where the model predictions have
been calculated using a \texttt{SARAH}-generated \texttt{SPheno}
version~\cite{Porod:2003um,Porod:2011nf,Staub:2008uz,Staub:2013tta}. In the
following discussion we mostly focus on the interplay between the dark matter
phenomenology and the strengths of the EWPT.
\begin{table}
\center
\begin{tabular}{|c|ccccc|ccc|c|}
\hline
 & $M_H$ & $M_A$ & $M_{H^\pm}$ & $\lambda_L$ & $\lambda_2$ & $T_c$ & $T_n$ & $v_c/T_c$ & $\mu_{\gamma \gamma}$\\
\hline 
BM1 & 66 & 300 & 300 & $1.07\times 10^{-2}$ & 0.01 & 113.3 & 110.3 & 1.5 & 0.90\\
BM2 & 200 & 400 & 400 & 0.01 & 0.01 & 116.1 & 113.7 & 1.5 & 0.93\\
BM3 & 5 & 265 & 265 & -$6\times10^{-3}$ & 0.01 & 118.2 & 116.3 & 1.3 & 0.90\\
\hline
\end{tabular}
\caption{Input parameters for the three benchmark scenarios discussed in the text along with 
critical and nucleation temperatures, the transition strength and the signal strength for $h\rightarrow \gamma\gamma$. 
The masses, given in $\GeV$, are pole masses and the couplings $\lambda_i$ are specified at 
$Q = M_Z$. Temperatures are also given in $\GeV$.}

\label{tab:benchmarks}
\end{table}

\begin{figure}
\center
\includegraphics{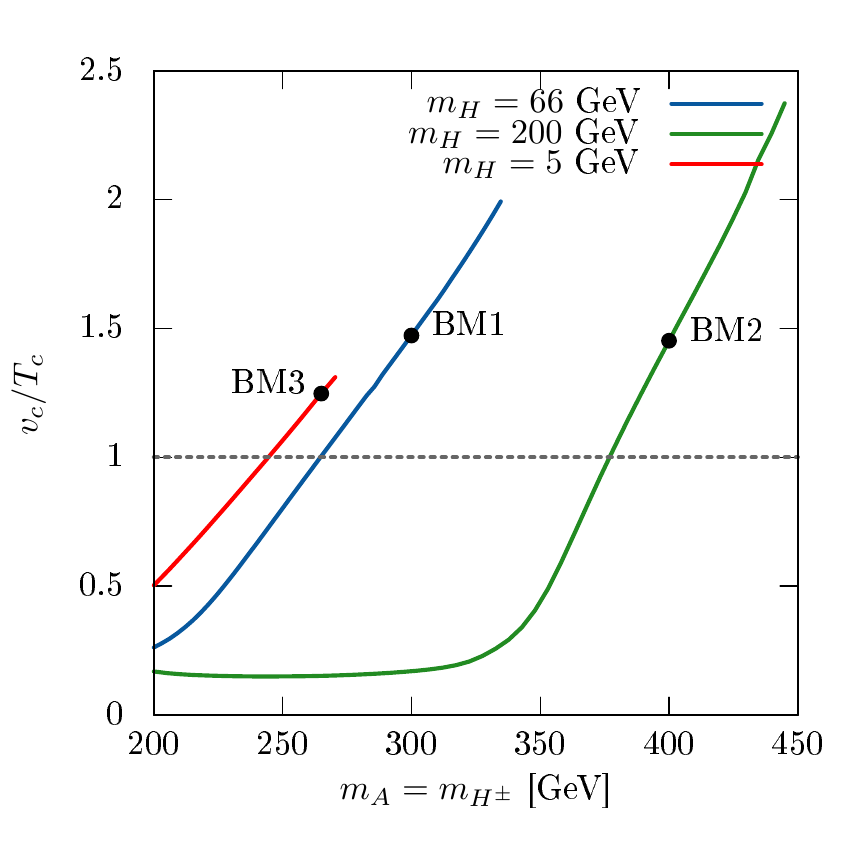}
\caption{Phase transition strength as a function of the heavier IDM scalar masses, taking $m_A = m_{H^\pm}$. The remaining parameters are 
chosen as in the benchmark models of Table~\ref{tab:benchmarks}, which are shown by black dots. The lines for BM1 and BM3 terminate where the inert doublet develops a non-zero vev, $\phi\neq0$, as described in the text. \label{fig:bm_vary}}
\end{figure}

Our key finding is that the requirement of a strongly first-order phase transition generally leads to 
a large mass splitting between the LOP and the other scalars in the IDM. 
Our benchmark models are summarized in Tab.~\ref{tab:benchmarks}, along with the corresponding critical and nucleation 
temperatures, as well as phase transition strengths, as parametrized by the ratio $v_c/T_c$. 
In each case the masses of the $A$ and $H^\pm$ 
are chosen to ensure a strongly first-order phase transition. In Fig.~\ref{fig:bm_vary} we show 
the dependence of the transition strength on these parameters. The lines corresponding 
to BM1 and BM3 terminate where the potential develops a non-inert ($\phi\neq 0$) vacuum 
first during thermal evolution. This vacuum can then either continuously evolve 
into the SM/inert ($\phi=0$) vacuum at $T=0$ or it can persist to low temperatures. In the latter case, the EWPT 
can occur in two steps. Such models are viable if the inert vacuum 
is deeper than the new one at $T=0$ and both transitions complete (\ie nucleation 
rate(s) are large enough). Two step electroweak phase transitions have 
been investigated in detail in Refs.~\cite{Patel:2012pi,Blinov:2015sna}. In this work 
we consider only simple one step transitions, hence the truncation. Notice that
in Ref.~\cite{Gil:2012ya} the strength of the EW phase transition in models
with multiple phase transition steps was always weaker, see Fig.~3 and 4 in Ref.~\cite{Gil:2012ya}.

First, let us consider model BM1, where LOP production in the early Universe is 
predominantly set by near-resonant $s$-channel Higgs exchange. 
This scenario has been recently examined in the context of phase transitions 
in Ref.~\cite{Gil:2012ya}. Even more recently, it has also been suggested as a
possible explanation~\cite{Modak:2015uda} of the Fermi-LAT gamma-ray excess
(see Ref.~\cite{Daylan:2014rsa} and references therein). 
As discussed above, the on-resonance requirement forces $m_H\sim m_h/2$, but
allows $\lambda_L$ to be small enough to be consistent with direct detection constraints.
Here DM production does not rely on interactions with 
$A$ or $H^\pm$, so their masses can be essentially chosen freely, as long as the resulting 
quartic couplings $\lambda_i$ (through Eq.~\eqref{eq:treemasses}) satisfy perturbativity and 
constraints from electroweak precision observables (EWPO), which we check with \texttt{2HDMC}~\cite{Eriksson:2009ws}. 
In order to satisfy the BNPC of Eq.~\eqref{eq:bnpc_gauge_variant} one needs to increase 
the coupling of the new scalars to $h$, which, in turn increases the splitting 
of $A$ and $H^\pm$ relative to $H$. For this and the following models we choose $m_A = m_{H^\pm}$ to minimize the impact of 
splitting these states from $H$ on the Peskin-Takeuchi $T$ function~\cite{Barbieri:2006dq} and to reduce the number 
of parameters. This assumption can be easily relaxed, but the results are qualitatively similar. 
This benchmark represents the only class of scenarios 
where the thermal LOP relic density (which we calculated with the
\texttt{micrOMEGAs} code~\cite{Belanger:2013oya}) matches the observed DM
universal density, and where the EW phase transition is strong-enough first
order.

When $m_A,\; m_{H^\pm} \gtrsim 340$ GeV, the corresponding loop corrections to $m_H$ are large 
and require $\mu_2^2<0$. This causes a second minimum to appear in the potential 
at $T=0$. As $m_A = m_{H^\pm}$ is increased further, this minimum quickly 
becomes deeper than the SM one, corresponding to the termination of 
the blue curve in Fig.~\ref{fig:bm_vary} at $m_A\sim 350\gev$. This behaviour was also observed in 
Ref.~\cite{Gil:2012ya}.

In this scenario, the LOP mass $m_H$ has been chosen slightly above the 
kinematic threshold of the decay $h\to HH$ in order to evade 
constraints from direct searches for invisible Higgs decays and Higgs rate measurements. 
These however become important for our benchmark scenario BM3 (see below).

The second benchmark BM2 in Tab.~\ref{tab:benchmarks} represents the intermediate mass regime. 
Here annihilation into gauge bosons is efficient and DM is generally underabundant, unless 
there is a cancellation among different amplitudes~\cite{Goudelis:2013uca}. 
The cancellation depends, as indicated above, on how close $\lambda_{L,S}\to 0$, \ie 
on how degenerate the IDM Higgs sector is. In our benchmark, such a cancellation requires 
$m_H\approx m_A \approx m_{H^\pm}$ with a maximum splitting of $\sim 10$ GeV. 
These small splittings 
lead to small couplings of the new states to $h$ and therefore an insufficiently strong 
phase transition. Thus the phase transition requirement forces \emph{thermal relic} DM 
to be underabundant. The observed DM density can be explained here, however, by invoking non-thermal 
production mechanisms (\eg the decay of a heavy particle) or with the existence of additional 
DM particles (\eg axions). The multitude of ``non-standard'' 
production mechanisms has been recently reviewed in Ref.~\cite{Baer:2014eja}. 

The final benchmark model, BM3, belongs to the light-mass regime, and is another example that requires further ingredients to be fully 
consistent with the phenomenology of the DM sector. For $m_H < m_h/2$, decays of the SM 
Higgs to invisible final states become possible, with a decay rate~\cite{Barbieri:2006dq} 
\beq
\Gamma (h\rightarrow HH) = \frac{v^2\lambda_L^2}{8\pi m_h}\left(
1-\frac{4m_H^2}{m_h^2}
\right)^{1/2}
.
\eeq 
Requiring consistency with the observed $95\%~\mathrm{C.L.}$ upper limit on the
branching fraction, $\mathrm{BR}(h\to HH) \le 17\%$~\cite{Bechtle:2014ewa},
provides a strong constraint on the coupling $\lambda_L$ of
$|\lambda_L|\lesssim 0.007$, while a large value $|\lambda_L|\gtrsim 0.4$ is
required to sufficiently deplete the DM abundance~\cite{Goudelis:2013uca}.
These problems can be remedied by softly breaking the $\Ztwo$, which would
allow $H$ to decay~\cite{Enberg:2013jba,Enberg:2013ara}. As in the previous
example, another explanation for DM is then needed. An alternative possibility
is to provide the LOP with new annihilation modes, \eg to new light vector
bosons~\cite{Ko:2014uka}, or a mechanism to dilute the thermal relic density,
such as an episode of late entropy injection~\cite{Gelmini:2006pq,
Wainwright:2009mq}.

DM phenomenology aside, it is again easy to get a strongly first-order phase transition 
with a large mass splitting between $H$ and $A$, $H^\pm$. We note that this scenario 
requires a significant tuning of parameters, because a small LOP mass requires near cancellation of 
tree-level and loop contributions. For $\lambda_L >0$, this leads to negative values 
of $\mu^2_2$ which can result in the appearance of a new $\phi\neq 0$ minimum.

 In all three cases, the first-order transition is driven by the non-analytic $(m^2)^{3/2}$ terms 
(see Eq.~\ref{eq:Jb_highT}) due to $A$ and $H^\pm$, while the gauge boson contributions 
are not as important. This explains the common feature of large splittings between $H$ and 
$A$, $H^\pm$ among the benchmark scenarios. 
These lead to large couplings between $h$ and the new states, enhancing the 
size of thermal corrections.  
This appears to be a generic requirement for increasing 
the strength of the phase transition in the IDM.
In particular, for the benchmark scenarios BM1, 2 and 3 we have 
$(\lambda_3,\;\lambda_4\;,\lambda_5)\approx(3.3, -1.7, -1.5)$, $(4.6,-2.3,-2.0)$ and $(2.7,-1.4,-1.2)$, 
respectively, at the scale $Q=246\;\GeV$. These coupling constants are large, but remain perturbative at 
energy scales of interest, which we checked using the RG equations in Appendix~\ref{sec:rge}.
It is also important to emphasize that 
thermal corrections to the crucial $(m^2)^{3/2}$ terms from $A$ and $H^\pm$ 
are not subject to gauge invariance issues that affect the gauge sector contributions. 
As a result, we expect these arguments to remain valid in the context of a fully gauge invariant treatment. 
This can be further tested in a toy model with all gauge coupling constants set to 0, thereby 
completely eliminating gauge dependence from the effective potential.\footnote{We thank Michael Ramsey-Musolf 
for pointing this out to us.} We checked that such a simplified analysis gives quantitatively similar 
results for critical temperatures and transition strengths when the scalar couplings are large.

The high-$T$ expansion of the effective potential also provides a simple
explanation for the shape of the curves in Fig.~\ref{fig:bm_vary}. In this
limit the transition strength is proportional to the coefficient of the $v^3$
term~\cite{Quiros:1999jp}. For the IDM scalars such terms arise from the
non-analytic contributions proportional to $(\mu^2_2 + \lambda_S v^2_c)^{3/2}$
(assuming $m_A = m_{H^\pm}$, as above, and ignoring daisy contributions for simplicity), 
which behaves as $v^3$ only when
$\lambda_S v^2_c \gg \mu_2^2$. Thus when $\lambda_S v^2_c \ll \mu_2^2$, the
transition strength is independent of IDM parameters, corresponding to the
plateau of the green curve in Fig.~\ref{fig:bm_vary}. In the opposite limit,
the IDM gives an additional contribution to the cubic coefficient, so the
transition strength scales as $v_c/T_c \sim \lambda_S^{3/2}\sim m_A^3$, as illustrated by
the monotonically increasing sections of the curves in Fig.~\ref{fig:bm_vary}.\footnote{The precise scaling is modified by Daisy corrections, $\mathcal{O}(\mu_2^2/\lambda_S v_c^2)$ 
terms, finite-$T$ and renormalization group effects.}

For heavy masses $m^2/T^2\gg 1$ (with $T\sim 100$ GeV), 
the IDM states thermally decouple, but this does not mean that they 
have no impact on the phase transition. When $\mu_2^2 \gg |\mu_1|^2$, 
the heavy doublet can be integrated out to yield a SM effective 
theory with the potential 
\beq
V_0 = \mu^2 |H|^2 + \lambda |H|^4 + \kappa |H|^6+\dots
\label{eq:treepot_eft}
\eeq
where the dots stand for higher mass dimension operators. 
The parameters $\mu^2$, $\lambda$ and $\kappa$ can be related 
to those in the fundamental IDM by equating the effective 
potentials for the two models at a \emph{matching} scale $Q\sim \mu_2$. 
For example, one-loop matching yields 
\beq
\kappa = \frac{1}{24\pi^2\mu_2^2}(\lambda_L^3 + \lambda_S^3 +\lambda_3^3/4),
\label{eq:kappa_eft}
\eeq
while $\mu^2$ and $\lambda$ are determined below the matching scale by 
fixing the VEV and the Higgs mass. 
With the presence of a dimension-six term in the potential, the barrier required for 
a strongly first-order transition can be generated 
if $\lambda\lesssim 0$ 
and $\mu^2 + c T^2>0$ for $T\sim T_c$, where $c$ encodes thermal corrections from 
SM states only~\cite{Carrington:1991hz}. Scenarios of this type have been considered, 
 \eg in Refs.~\cite{Grojean:2004xa,Delaunay:2007wb,Grinstein:2008qi,Chung:2012vg}.
One immediate difficulty is that in the IDM $\kappa$ is generated only at one-loop, so 
in order for this operator to be significant for field values of around the electroweak VEV, 
one must overcome the loop suppression, suggesting that the combination 
$\lambda_L^3 + \lambda_S^3 +\lambda_3^3/4$ cannot be too small. This again forces a 
large splitting between the IDM states, meaning that the heavy DM scenario described in 
Section~\ref{sec:idm_dm} cannot be realized together with a strongly first-order phase transition. 
Such large couplings can run into perturbativity problems and invalidate the expansion 
used to generate the effective field theory.

We briefly comment on the discovery prospects of the new IDM states at
the LHC. Due to the $\Ztwo$ symmetry, the IDM states can only be produced
pairwise at colliders. Successive decays of the heavier IDM states $A$ and
$H^\pm$ into the LOP and a $Z$ or $W$ boson, respectively, can give rise to
multilepton signatures~\cite{Miao:2010rg,Gustafsson:2012aj,Belanger:2015kga}.
In a recent analysis~\cite{Belanger:2015kga} of LHC searches for supersymmetric
particles with two leptons plus missing transverse energy in the final state in
the context of the IDM, mass limits of up to $m_A \lesssim 140~\GeV$ for LOP
masses $m_H \lesssim 55~\GeV$ and charged Higgs masses around $85 - 150~\GeV$
have been derived. While these limits partly exceed previous limits from the
LEP collider, they are not yet sensitive to the parameter regions that yield a
strongly first-order phase transition as required for successful electroweak
baryogenesis, see Fig.~\ref{fig:bm_vary}. 

The Higgs portal coupling $\lambda_L$ 
can be probed directly at the LHC by searching for, \eg jets plus missing 
energy~\cite{Craig:2014lda}. However, the small magnitude of $\lambda_L$ in our 
benchmarks (as required by direct detection or relic abundance constraints) suggests 
that these searches will not be sensitive to the models in Tab.~\ref{tab:benchmarks}.

The new IDM states can also have an indirect effect on precision Higgs measurements. 
In particular, the new charged state $H^\pm$ provides an additional contribution 
to the loop-induced $h\rightarrow \gamma\gamma$ and $\gamma Z$ rates. These 
effects have been recently studied in Refs.~\cite{Swiezewska:2012eh,Arhrib:2012ia} in 
the context of the 125 GeV Higgs boson. 
Modifications of these branching fractions by $\mathcal{O}(10\%)$ are a {\em generic} 
feature of our benchmark scenarios, as we show below.  The $h\rightarrow\gamma\gamma$ 
rate has the form~\cite{hhg,Branco:2011iw,Djouadi:2005gj,Spira:1995rr,Swiezewska:2012eh,Arhrib:2012ia}
\beq
\Gamma (h\rightarrow \gamma\gamma)
= \frac{\alpha^2  G_F m_h^3}{128\sqrt{2}\pi^3}\left|
\mathcal{A}_{\rm SM} + \frac{\lambda_3 v^2}{2m^2_{H^\pm}} A_0\left(\frac{m_h^2}{4m^2_{H^\pm}}\right)
\right|^2,
\eeq
where the leading contributions to the SM amplitude $\mathcal{A}_{\rm SM} \approx -6.56 + 0.08i$ come 
from $W$ bosons and top quarks. The second term is the new contribution from $H^\pm$, where 
$A_0$ is a loop function with the property $\lim_{x\rightarrow 0} A_0(x) = 1/3$~\cite{Spira:1995rr}. 
For our benchmarks we have $\lambda_3 > 0$, so one expects a suppression of $h\rightarrow \gamma\gamma$ relative to the 
SM.\footnote{Various limits on LOP-Higgs coupling discussed above force $|\lambda_L|$ to be small, such that 
the $H$ mass is primarily determined by $\mu_2^2$ (at tree level, see Eq.~\eqref{eq:treemasses}). If the charged Higgs $H^\pm$ 
is heavier than $H$ then this forces $\lambda_3>0$.} Note that for fixed $\mu_2^2$, the amplitude for the $H^\pm$ contribution tends to a 
constant value $1/3$ as $\lambda_3$ is increased. This means that in the limit of a large mass splitting between $H$ and $H^\pm$, which 
is required for a strongly first order phase transition, the branching fraction is reduced by $\sim 10\%$. 
This effect was also noticed in Refs.~\cite{Borah:2012pu,Cline:2013bln} for models similar to our BM1 and BM2, respectively.  
The deviations to $\mathrm{BR}(h\rightarrow \gamma\gamma)$ induced by $H^\pm$
are shown in Tab.~\ref{tab:benchmarks} in terms of the SM normalized signal 
strength $\mu_{\gamma\gamma} = (\sigma\; \mathrm{BR})/(\sigma\; \mathrm{BR})_\mathrm{SM}$. 
While they are still consistent with the present measurements from ATLAS~\cite{Aad:2014eha} and CMS~\cite{Khachatryan:2014ira},
the LHC should reach a precision of  4--8\% for $\mu_{\gamma\gamma}$~\cite{Bechtle:2014ewa,Dawson:2013bba}, thereby definitively testing 
the benchmark scenarios in Tab.~\ref{tab:benchmarks}. 

While our benchmarks feature sizable deviations of ${\rm BR}(h\rightarrow\gamma\gamma)$ from the SM expectation, 
we note that it {\em is} possible to avoid this by taking $H^\pm$ to be nearly degenerate 
with $H$, and using $A$ alone to drive the phase transition to be strongly first order. 
However, in this case, efficient coannihilation of $H$ with $H^\pm$ during freeze-out generally results in a 
very small relic abundance~\cite{Pierce:2007ut}. The near degeneracy is also required by constraints on the oblique $T$ parameter 
when $m_A \gg m_{H^\pm}$~\cite{Barbieri:2006dq}. For example, taking $m_{H^\pm} = 70\gev$, $m_A = 370\gev$ and 
other parameters as in BM1 results in a strongly first order phase transition, an order of magnitude 
smaller relic abundance and only a $\sim 3\%$ depression of $\mu_{\gamma\gamma}$ relative to the SM.

\section{Discussion and Conclusions\label{sec:conclusion}}
We studied the structure of the electroweak phase transition in the inert Higgs
doublet model, utilizing a set of three benchmark scenarios that feature a potentially
viable dark matter particle. Our choices for the three benchmark models
essentially exhaust all possible prototypical setups for particle dark matter
in the inert doublet model. While only one of the benchmarks has a dark
matter particle with a thermal relic density matching the observed dark matter density, 
the other two (under- and over-abundant) can
be made viable by invoking additional production mechanisms or a
scenario where the thermal relic density is diluted away, respectively.

The key finding of our study is that in all cases where the model possesses a
reasonable particle dark matter candidate, the inert scalar spectrum can be
arranged in such a way so as to produce a strongly first-order electroweak
phase transition. Central to achieving such a phase transition is to postulate
a large enough splitting between the dark matter candidate and the heavier
inert scalars. The physics driving this result is simple: Large mass splittings
generically correspond to large couplings between the inert scalars and the
Standard Model-like Higgs; These, in turn, increase the magnitude of non-analytic $\sim (m^2)^{3/2}$
terms in the temperature-dependent effective potential and thus the potential
barrier between the field origin and the $SU(2)\times U(1)$-breaking phase. 
It is clear that this physical effect is generic, and, in fact, it has also been observed 
in other implementations of Two Higgs Doublet models~\cite{Cline:2011mm,Dorsch:2013wja,Fuyuto:2015jha}.

The mass splitting under consideration cannot be arbitrarily large. For large
enough values, for example, the phase structure of the model becomes more complicated, with
possible non-zero vacuum expectation values for the inert doublet and
multiple-step phase transitions. While, based upon the results of
Ref.~\cite{Gil:2012ya} the latter possibility is not expected to yield stronger
electroweak phase transitions than in the single-step case, this is an
interesting possibility which we leave for future studies.

The question of how to embed large-enough $CP$ violating sources in detail was
also left unanswered here. It will be interesting to study whether such a source
(for example an additional gauge-singlet complex scalar, see
Ref.~\cite{Bonilla:2014xba}) significantly impacts the
electroweak phase transition and dark matter phenomenology,
and, with this, the conclusions reached in the present study.
\section{Acknowledgements}
We thank Patrick Draper, Bj\"orn Herrmann, Jonathan Kozaczuk, Carlos Tamarit
and Florian Staub for useful discussions. We are grateful to Michael Ramsey-Musolf and David Morrissey for 
insightful comments about the manuscript. NB is supported by the National
Sciences and Engineering Research Council of Canada (NSERC). 
SP and TS are partly supported by the US Department of Energy, under Contract No. DE-FG02-04ER41286.
TS is furthermore supported by a Feodor-Lynen research fellowship sponsored by
the Alexander von Humboldt foundation.

\appendix
\section{Renormalization Group Equations\label{sec:rge}}
Here we list the beta functions for the inert 2HDM at one-loop order. 
The general form of the RG equations is 
\beq
\frac{d\lambda}{dt} = \frac{1}{16\pi^2} \beta_\lambda,
\eeq
where $t = \ln Q/Q_0$ and $Q_0$ is a reference scale. We take $Q_0 = M_Z$. 
The $U(1)_Y$ gauge coupling has the GUT normalization: $g_1 = \sqrt{5/3} g'$. 
The beta functions below have been checked with \sarah~\cite{Staub:2013tta}.
Partial one 
loop results can be found in Refs.~\cite{Ferreira:2009jb,Branco:2011iw, Goudelis:2013uca} 
for dimensionless parameters only. These agree with the 
formulae below. 

The gauge coupling evolution is determined by 
\begin{align}
\beta_{g_1} & = \frac{21}{5} g_1^3\\
\beta_{g_2} & = -3 g_2^3 \\
\beta_{g_3} & = -7 g_3^3.
\end{align}

For the third generation Yukawas we have
\begin{align}
\beta_{y_t} & = -\frac{17}{20} g_1^2 y_t-\frac{9}{4} g_2^2 y_t
-8 g_3^2 y_t +\frac{9 }{2} y_t^3 +\frac{3}{2} y_b^2 y_t + y_t y_{\tau }^2\\ 
\beta_{y_b} & = -\frac{1}{4} g_1^2 y_b-\frac{9}{4} g_2^2 y_b-8 g_3^2 y_b+\frac{3}{2} 
y_b y_t^2+y_b y_{\tau }^2+\frac{9}{2} y_b^3\\
\beta_{y_\tau} & = -\frac{9}{4} g_1^2 y_{\tau }-\frac{9}{4} g_2^2 y_{\tau }
+3 y_t^2 y_{\tau }+3 y_b^2 y_{\tau }+ \frac{5 }{2}y_{\tau }^3.
\end{align}

Next we consider the scalar potential parameters. The evolution of the dimensionless 
quartic couplings $\lambda_i$ is governed by

\begin{align}
\beta_{\lambda_1} & =-\frac{9}{5} g_1^2 \lambda _1-9 g_2^2 \lambda _1+\frac{27 }{200}g_1^4+\frac{9}{20} g_2^2 g_1^2\\ 
 & +\frac{9 }{8}g_2^4+24 \lambda _1^2+2 \lambda _3^2+\lambda _4^2+\lambda _5^2+2 \lambda _3 \lambda _4\nonumber\\ 
 & +12 \lambda _1 y_t^2-6 y_t^4+12 \lambda _1 y_b^2-6 y_b^4 + 4 \lambda _1 y_{\tau }^2-2 y_{\tau }^4\nonumber\\
\beta_{\lambda_2} & =-\frac{9}{5} g_1^2 \lambda _2-9 g_2^2 \lambda _2+\frac{27 }{200}g_1^4+\frac{9}{20} g_2^2 g_1^2+\frac{9 }{8}g_2^4+24 \lambda _2^2\\ 
 & +2 \lambda _3^2+\lambda _4^2+\lambda _5^2+2 \lambda _3 \lambda _4\nonumber\\
\beta_{\lambda_3} & =-\frac{9}{5} g_1^2 \lambda _3-9 g_2^2 \lambda _3+\frac{27 }{100}g_1^4-\frac{9}{10} g_2^2 g_1^2+\frac{9 }{4}g_2^4\\ 
 & +4 \lambda _3^2+2 \lambda _4^2+2 \lambda _5^2+12 \lambda _1 \lambda _3+12 \lambda _2 \lambda _3+4 \lambda _1 \lambda _4\nonumber\\ 
 & +4 \lambda _2 \lambda _4+6 \lambda _3 y_t^2+6 \lambda _3 y_b^2 +2 \lambda _3 y_{\tau }^2\nonumber\\
\beta_{\lambda_4} & =-\frac{9}{5} g_1^2 \lambda _4-9 g_2^2 \lambda _4+\frac{9}{5} g_2^2 g_1^2+4 \lambda _4^2+8 \lambda _5^2\\ 
 & +4 \lambda _1 \lambda _4+4 \lambda _2 \lambda _4+8 \lambda _3 \lambda _4+6 \lambda _4 y_t^2+6 \lambda _4 y_b^2 +2 \lambda _4 y_{\tau }^2\nonumber\\
\beta_{\lambda_5} & =-\frac{9}{5} g_1^2 \lambda_5-9 g_2^2 \lambda _5 + 4\lambda_1\lambda_5 +4\lambda_2\lambda_5 +8\lambda_3\lambda_5 
\\
&+12 \lambda_4\lambda_5 +6 \lambda_5 y_t^2 + 6\lambda_5 y_b^2 + 2 \lambda_5 y_\tau^2\nonumber
\end{align}

The beta functions for the mass parameters are given by 
\begin{align}
\beta_{\mu_1^2} & = -\frac{9}{10} g_1^2 \mu _1^2-\frac{9}{2} g_2^2 \mu _1^2
+12 \lambda _1 \mu _1^2+4 \lambda _3 \mu _2^2+2 \lambda _4 \mu _2^2\label{eq:mu12_beta1} \\ 
 & +6 \mu _1^2 y_t^2+6 \mu _1^2 y_b^2 + 2 \mu _1^2 y_{\tau }^2\nonumber\\
\beta_{\mu_2^2} & =-\frac{9}{10} g_1^2 \mu _2^2-\frac{9}{2} g_2^2 \mu _2^2
+12 \lambda _2 \mu _2^2+4 \lambda _3 \mu _1^2+2 \lambda _4 \mu _1^2.\label{eq:mu22_beta1}
\end{align}

Finally, the anomalous dimensions for the Higgs and the inert scalar are 
\begin{align}
\gamma_h & =-\frac{9}{20} g_1^2-\frac{9 }{4}g_2^2+3 y_t^2+3 y_b^2+ y_{\tau }^2\\
\gamma_\phi & =-\frac{9 }{20}g_1^2-\frac{9 }{4}g_2^2. 
\end{align}

\bibliographystyle{JHEP}
\bibliography{biblio}
\end{document}